*Ab Initio* calculation of the vibrational spectrum and thermodynamic properties of rhombohedral $P_4O_{10}$

James R. Rustad, Sullivan Park Research Center, Corning, Inc., Corning, NY 14831

**Abstract**—Plane-wave pseudopotential methods and density functional perturbation theory are used to calculate the phonon density of states and thermodynamic functions of $h$-$P_4O_{10}$. The calculated vibrational spectrum is in good agreement with the measured spectrum, but the calculations indicate some modifications in the interpretation of the spectrum, mainly suggesting changes in the number of components used to fit a few of the observed peaks. The calculated low-temperature heat capacity is in good agreement with the measured heat capacity, with a systematic offset of approximately -5 J mol$^{-1}$ K$^{-1}$, independent of temperature. . Comparison with molecular calculations indicate that molecular contributions make up about 80% of the heat capacity of $h$-$P_4O_{10}$

**Introduction**

Hexagonal tetraphosphorous decaoxide $h$-$P_4O_{10}$ is a molecular solid consisting of discrete $P_4O_{10}$ molecules. This phase is metastable with respect to orthorhombic polymorphs[1], but is the most common form of phosphorous oxide and is used as a reference in calorimetric studies of phosphate phases[2]. This paper reports the results of first-principles calculations of the phonon density of states, heat capacity and thermodynamic functions of $h$-$P_4O_{10}$ from 0-298.15 K   Low-temperature heat capacity data for $h$-$P_4O_{10}$ have been reported[3], and Raman and infrared vibrational spectra have been measured[4] for this system, making it an ideal system for comparing theoretical and experimental determination of thermodynamic properties. In this sense it can serve as a guide for *ab initio* calculation of the thermodynamic properties of phosphate oxide compounds. Moreover, it is interesting to have a quantitative understanding of the consequences of leaving out van der Waals forces in the thermodynamic properties of molecular solids.

**Methods**

The CASTEP code[5], as implemented within Materials Studio of Accelrys, Inc. was used for the electronic structure calculations. A plane-wave cutoff of 750 eV and norm-conserving (NC) pseudopotentials[6,7] and the PBE exchange-correlation functional[8] was used to do the calculations. The atom positions were optimized at a fixed cell length, starting from the measured structure[9]. Forces were optimized with the BFGS algorithm to a tolerance of 0.003 eV/Å. The cell length was fixed because the molecular forces which hold the crystal together are not well-described by density functional theory of electronic structure without empirical corrections. The phonons were calculated with six **q**-points {(1/3, 1/3, 1/3), (1/3, 1/3, 0), (1/3, 1/3,-1/3), (1/3, 0, 0), (1/3, 0, -1/3), (0,0,0)} using the linear-response method[10].

**Results and Discussion**

*Vibrational Spectrum*

Vibrational frequencies at **q**=(0, 0, 0) are given in Table 1. If the calculated frequencies are scaled by ~1.025, a good match is achieved with most of the measured frequencies. Several revisions to the interpretation of the measured spectrum are suggested by the calculations. The relevant regions are highlighted in bold italics in Table 1.

(1) The observed peak at 277 cm$^{-1}$ has three components, not two, with 2 E modes and an $A_1$ mode
(2) The peak at 292 cm$^{-1}$ is not a combination mode (or at least has a contribution from fundamentals )
(3) The peak at 425 cm$^{-1}$ consists of six components (2 $A_1$ and 4 E), not three.
(4) The region in the partly resolved $A_1$ peak at 578 cm$^{-1}$ and the E peak at 579 cm$^{-1}$ have a contribution from an additional E vibration.
(5) The peaks observed at 708 cm$^{-1}$ and 720 cm$^{-1}$ may reflect coupling of near degenerate active and inactive bands calculated at 692 cm$^{-1}$.
(6) The region near the two peaks observed at 829 cm$^{-1}$ and 846 cm$^{-1}$ have contributions from 5 modes (one $A_1$ and four E)
(7) The band at 1036 cm$^{-1}$ previously assigned to a (762cm$^{-1}$+277cm$^{-1}$) combination may be a fundamental (calculated at 1006 cm$^{-1}$).

The most problematic region is at high frequency. In the measured spectrum, there are two peaks in this region: the main peak at 1419 cm$^{-1}$ and a second peak near 1386 cm$^{-1}$. In the calculations two peaks are also evident, red-shifted relative to the measurements, at 1358 cm$^{-1}$ and 1387 cm$^{-1}$. In the measured spectrum, the high frequency peak is the most complex, with a shoulder at 1417 cm$^{-1}$, along with a partly resolved peak at 1413 cm$^{-1}$. The peak at 1386 cm$^{-1}$ appears to be a single-component peak. In contrast, the calculations give a single peak at 1387 cm$^{-1}$, but a three component peak (with 1 $A_1$ and 2 E components) between 1356 cm$^{-1}$-1359 cm$^{-1}$. The calculated phonon density of states is given in Figure 1.

For comparison with CASTEP, the vibrational frequencies of $P_4O_{10}(g)$ were calculated at the PBE/aug-cc-pVTZ[10] and B3LYP[11,12]/aug-cc-pVTZ level using Gaussian09[13]. The results are given in Table 2. For the same PBE functional, the tendency to underestimate vibrational frequencies is much greater in the Gaussian orbital calculations than for the plane-wave calculations. For the gaussian orbital calculations, frequencies need to be scaled, on average, by approximately 1.08 to match measured frequencies. For the molecular calculations, B3LYP gives a much closer match to the experimental frequencies, requiring an average scaling factor of approximately 1.03.

*Thermodynamic Properties and the Zero-Point Energy*

The heat capacity is computed from the phonon density of states $g(\omega)$ using:

$$C_v(T) = R \int \frac{\left(\frac{\hbar\omega}{kT}\right)^2 \exp\left(\frac{-\hbar\omega}{kT}\right)}{\left[1 - \exp\left(\frac{-\hbar\omega}{kT}\right)\right]^2} g(\omega) d\omega \qquad (1)$$

Figure 2 shows the good agreement between the calculated $C_v$ and the measured constant-pressure heat capacity[3], the calculated $C_v$ showing a systematic ~-5 J mol$^{-1}$ K$^{-1}$ offset relative to the measured $C_p$. This offset is nearly independent of temperature. Vibrational thermodynamic functions at 298.15 K are $C_v$(298.15 K)= 206.2 J K$^{-1}$ mol$^{-1}$; H(298.15)-H(0)=32.74 kJ mol$^{-1}$; S(298.15)= 212.9 J K$^{-1}$ mol$^{-1}$. The corresponding measured thermodynamic functions in Ref. 3 are $C_p$(298.15 K)=211.71 J K$^{-1}$ mol$^{-1}$, H(298.15)-H(0)=33.96 kJ mol$^{-1}$, S(298.15)= 228.86 J K$^{-1}$ mol$^{-1}$. The heat capacity as a function of temperature is given in Table S1 in the supporting information. The vibrational contributions to the enthalpy $\int C_v dT$, entropy $\int C_v/T dT$ and free energy (H-TS) are given in Table S2

The calculated zero point energy of $h$-$P_4O_{10}$ is given in Table 3 along with calculations on $P_4O_{10}(g)$ at various levels of theory. For $P_4O_{10}(g)$, zero point energies were calculated using the PBE and B3LYP functionals at both the 6-31G* and aug-cc-pVTZ basis sets with Gaussian09.

Table 2 allows a rough estimate of the corrections to the thermodynamic functions at the B3LYP level. B3LYP calculations could not be carried out within CASTEP because of the expense of hybrid functionals when used in plane-wave codes. Correcting the PBE frequencies by the average B3LYP/PBE scaling factor of 1.052, gives the following estimates for values of thermodynamic functions at the hybrid B3LYP level: $C_v$(298.15 K)= 199.2 J K$^{-1}$ mol$^{-1}$, H(298.15)-H(0)=31.2 kJ mol$^{-1}$, S(298.15)= 211.8 J K$^{-1}$ mol$^{-1}$. The estimated correction to the heat capacity is small, and in the wrong direction relative to experiment indicating that a more expensive treatment of the plane-wave calculation with hybrid functionals would not help to correct the systematic offset in the calculated and measured heat capacities.

Accounting for van der Waals forces would be a good direction for theoretical improvement, however, given the good agreement between theory and experiment the vibrational and thermodynamic contributions of the van der Waals interactions in this particular molecular solid appear to be small. It is interesting to note that the $P_4O_{10}$(g) molecule serves as a poor surrogate for the solid. Molecular vibrational heat capacities are 168.92 and 161.52 J mol-1 K-1 for PBE/aug-cc-pVTZ and B3LYP aug-cc-pVTZ, respectively. The hindered rotations of the $P_4O_{10}$ molecules evidently account for about 20% of the heat capacity in $h$-$P_4O_{10}$.

**Conclusions**

This paper reports a first-principles calculation of the phonon density of states and vibrational thermodynamic of $h$-$P_4O_{10}$ through calculation using density functional theory. The calculated frequencies agree well with spectroscopic measurements[4], but suggest some modification in the interpretation and fitting of the spectrum. The calculated heat capacity is in good agreement with experimental measurements[3], but show a consistent offset of approximately -5 J mol$^{-1}$ K$^{-1}$, independent of temperature. Estimated corrections to the thermodynamic functions at the B3LYP level are small and in the opposite direction relative to what would be required to account for the discrepancy with experiment. Comparison with molecular calculations indicate that molecular contributions make up about 80% of the heat capacity of $h$-$P_4O_{10}$

**Supporting Information**

Table S1, calculated heat capacity of $h$-$P_4O_{10}$, Table S2, calculated thermodynamic functions of $h$-$P_4O_{10}$. Table S3. Calculated coordinates for $h$-$P_4O_{10}$ in crystallographic information file format.

**References**

(1) Greenwood, N. N.; Earnshaw A. *Chemistry of the Elements*, Butterworth Heinemann, Oxford, 1985.
(2) Ushakov, S.V.; Helean, K.B.; Navrotsky, A. *J. Mater. Res.* **2002**, 16, 2623.
(3) Andon, R.J.L.;Counsell, J. F.; McKerrell, H.; Martin, J. F. *Trans. Faraday Soc.*, **1963**, 59, 2702
(4) Gilliam, S.J.; Kirby, S.J.; Merrow, C.N.; Zeroka, D.; Banerjee, A.; Jensen, J.O. *J. Phys. Chem. B*. **2003**, *107*, 2892.
(5) Clark, S.J.; Segall, M.D.; Pickard, C.J.; Hasnip, P.J.; Probert, M.J.; Refson, K.; Payne, M.C. *Zeitschrift fuer Kristallographie*, **2005**, *220*, 567.
(6) Lee, M.H. Ph.D. Thesis, Cambridge University, **1996**.
(7) Lin, J. S.; Qteish, A.; Payne, M. C.; Heine, V., *Phys. Rev. B*, **1993**, *47*, 4174.
(8) Perdew, J.P.; Burke, K.; Ernzerhof, M. *Phys. Rev. Lett*. **1996**, *77*, 3865.
(9) Cruickshank, D.W.J. Acta Cryst **1964**, *17*, 677.
(10) Refson, K.;Clark, S.J.; Tulip, P.R. Phys. Rev. B **2006**, *73*, 155114.
(11) Kendall, R.A.; Dunning, T.H.; Harrison, R.J. *J. Chem. Phy*s. **1992**, *96*, 6796.
(12) Becke, A.D. *J. Chem. Phys.* 1993, 98, 5648.
(13) Lee, C.T.; Yang, W.T.; Parr, R.G., Phys. Rev. B, 37, 785-789.
(14) Gaussian 09 , Revision A.1, Frisch, M.J. et al. Gaussian, Inc., Wallingford CT, 2009.

**Table 1.** Calculated and Measured Vibrational Frequencies of $h$-$P_4O_{10}$.

| $\omega(cm^{-1})$ | measured[a] | | $\omega(cm^{-1})$ | measured[a] |
|---|---|---|---|---|
| 50.21 | | | 544.22 | * |
| 50.21 | | | 545.31 | 560 |
| 65.41 | | | **570.35** | **578** |
| 70.57 | | | **570.43** | * |
| 75.13 | | | **570.43** | * |
| 75.13 | | | **572.40** | **579** |
| 87.66 | | | **572.40** | **579** |
| 87.66 | | | 573.55 | * |
| 96.31 | | | **692.16** | **708** |
| 258.70 | 257 | | **692.64** | **720** |
| 258.70 | 257 | | 740.39 | 762 |
| 262.40 | 262 | | 742.25 | 772 |
| 262.40 | 262 | | 742.25 | 772 |
| **278.35** | * | | 750.37 | * |
| **278.35** | * | | 752.78 | 786 |
| **283.70** | **277** | | 752.78 | 786 |
| **284.98** | | | **811.85** | **829** |
| **285.40** | **284** | | **811.85** | **829** |
| **285.40** | **284** | | **824.11** | **846** |
| 292.65 | | | **824.11** | **846** |
| 294.44 | 292? | | **824.92** | * |
| 294.44 | 292? | | **824.92** | * |
| 304.45 | * | | **829.50** | * |
| 304.45 | * | | **829.50** | * |
| 308.83 | * | | **830.67** | * |
| 326.09 | * | | 830.93 | * |
| 326.09 | * | | 973.25 | 1004 |
| 331.75 | 329 | | 977.63 | 1008 |
| 331.75 | 329 | | 977.63 | 1008 |
| **412.52** | * | | 999.11 | |
| **412.57** | **419** | | 1006.27 | 1036 comb? |
| **412.57** | **419** | | 1006.27 | 1036 com? |
| **413.06** | * | | 1356.22 | 1386 |
| **416.08** | * | | 1356.97 | 1386 |
| **416.08** | * | | 1356.97 | 1386 |
| **420.94** | **425** | | 1358.97 | 1413? |
| **422.14** | * | | 1358.97 | 1417? |
| **422.14** | * | | 1359.53 | * |
| **425.45** | **429** | | 1387.45 | 1419 |
| **425.45** | **429** | | 1404.34 | * |
| 428.89 | * | | | |

[a]Ref. 4
*Asterisks indicate missing modes in fitting the measured spectrum in Ref. 4

**Table 2.** Calculated frequencies for $P_4O_{10}(g)$*

| PBE | B3LYP | exp Raman | exp (IR) | degeneracy | Raman | IR | scaling B3LYP/PBE |
|---|---|---|---|---|---|---|---|
| 234 | 242 | 254 | | 2 | Y | N | 1.034 |
| 242 | 256 | | | 3 | N | N | 1.058 |
| 252 | 262 | 264 | 270 | 3 | Y | Y | 1.040 |
| 306 | 323 | | | 2 | Y | N | 1.056 |
| 376 | 399 | 411 | 409 | 3 | Y | Y | 1.061 |
| 379 | 401 | | | 3 | N | N | 1.058 |
| 508 | 535 | 553 | | 1 | Y | N | 1.053 |
| 518 | 559 | | 575 | 3 | Y | Y | 1.079 |
| 650 | 697 | 717 | | 1 | Y | N | 1.072 |
| 710 | 741 | | 763 | 3 | Y | Y | 1.044 |
| 767 | 813 | | | 3 | N | N | 1.060 |
| 780 | 805 | | | 2 | Y | N | 1.032 |
| 939 | 994 | | 1012 | 3 | Y | Y | 1.059 |
| 1342 | 1394 | 1406 | 1406 | 3 | Y | Y | 1.039 |
| 1371 | 1425 | 1440 | | 1 | Y | N | 1.039 |

*$T_d$ symmetry, Basis set aug-cc-pVTZ

**Table 3.** Zero-point energies for $P_4O_{10}(g)$ and $P_4O_{10}(s)$ (kJ/mol)

|  | $P_4O_{10}(g)$ | $P_4O_{10}(s)$ |
|---|---|---|
| CASTEP/NC/750 eV |  | 141.30 |
| B3LYP/6-31G* | 134.72 |  |
| B3LYP/aug-cc-pVTZ | 136.77 |  |
| PBE/6-31G* | 128.17 |  |
| PBE/aug-cc-pVTZ | 130.05 |  |

**Figure Captions**

Figure 1. Calculated phonon density of states g($\omega$) for $h$-$P_4O_{10}$. Instrument broadening is 0.05 THz, normalized to unity.

Figure 2. Calculated ($C_v$) and measured ($C_p$) heat capacities for $h$-$P_4O_{10}$

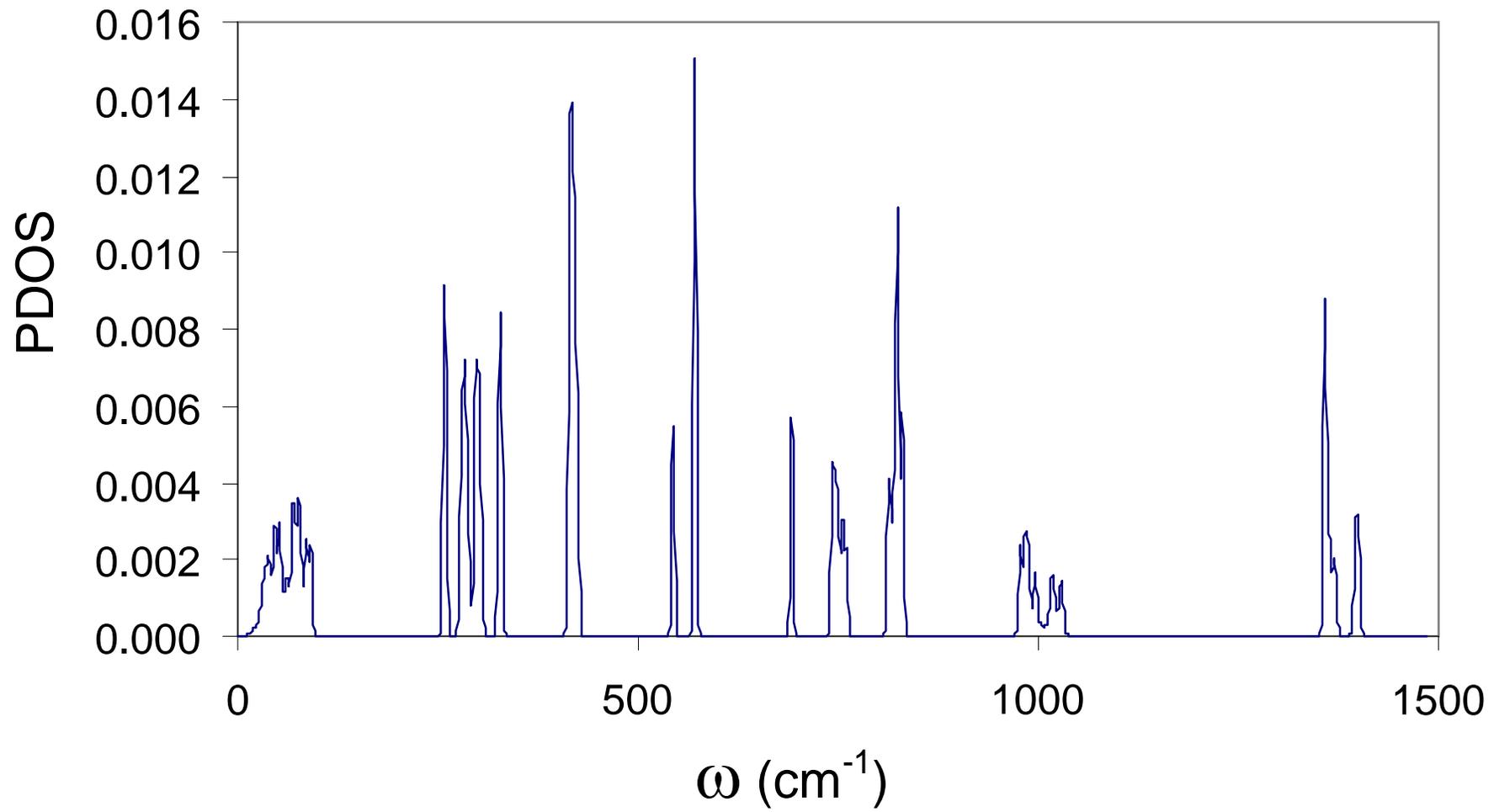

Figure 1

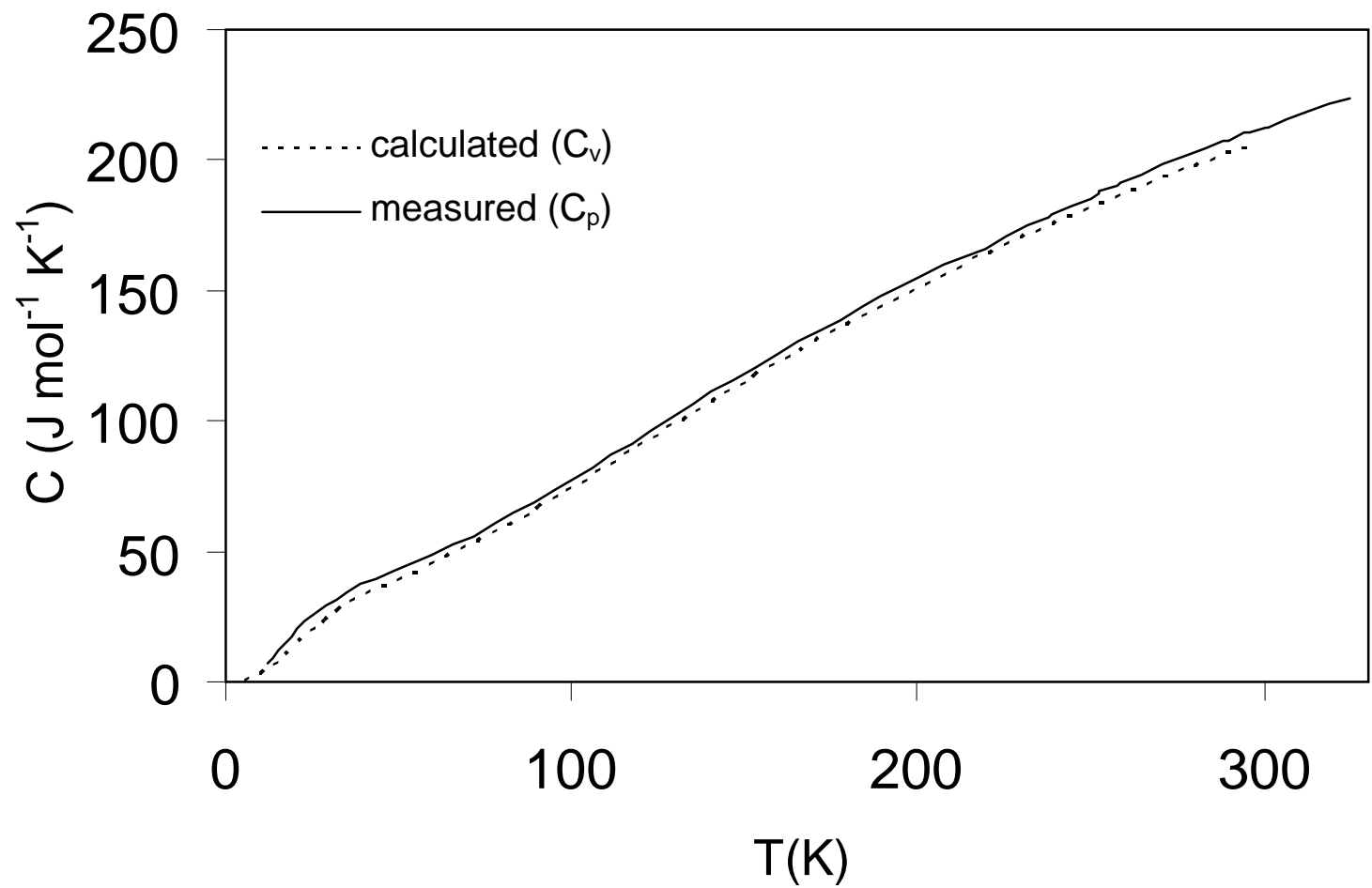

Figure 2